# Inducing Metallicity in Graphene Nanoribbons via Zero-Mode Superlattices

Daniel J. Rizzo[1]†, Gregory Veber[2]†, Jingwei Jiang[1,3]†, Ryan McCurdy[2], Ting Cao[1,3], Christopher Bronner[1], Ting Chen[1], Steven G. Louie[1,3]*, Felix R. Fischer[2,3,4]*, Michael F. Crommie[1,3,4]*

[1]Department of Physics, University of California, Berkeley, CA 94720, USA.

[2]Department of Chemistry, University of California, Berkeley, CA 94720, USA.

[3]Materials Sciences Division, Lawrence Berkeley National Laboratory, Berkeley, CA 94720, USA.

[4]Kavli Energy NanoSciences Institute at the University of California Berkeley and the Lawrence Berkeley National Laboratory, Berkeley, California 94720, USA.

*Correspondence to: crommie@berkeley.edu, ffischer@berkeley.edu, sglouie@berkeley.edu

†These authors contributed equally.

**Abstract**

The design and fabrication of robust metallic states in graphene nanoribbons (GNRs) is a significant challenge since lateral quantum confinement and many-electron interactions tend to induce electronic band gaps when graphene is patterned at nanometer length scales. Recent developments in bottom-up synthesis have enabled the design and characterization of atomically-precise GNRs, but strategies for realizing GNR metallicity have been elusive. Here we demonstrate a general technique for inducing metallicity in GNRs by inserting a symmetric superlattice of zero-energy modes into otherwise semiconducting GNRs. We verify the resulting metallicity using scanning tunneling spectroscopy as well as first-principles density-functional theory and tight binding calculations. Our results reveal that the metallic bandwidth in GNRs can be tuned over a wide range by controlling the overlap of zero-mode wavefunctions through intentional sublattice symmetry-breaking.

Extended two-dimensional (2D) graphene is renowned for being a gapless semimetal, yet when it is laterally confined to nanometer-scale one-dimensional (1D) ribbons a sizable energy gap emerges (*1, 2*). Unlike carbon nanotubes (which can exhibit metallicity depending on their chirality), isolated armchair and zigzag graphene nanoribbons (GNRs) always feature a band gap that scales inversely with the width of the ribbon (*2*). This is a selling point for GNRs since it makes them attractive as transistor elements for logic devices at the ultimate limits of scalability (*11*). But it is also a limitation since metallic GNRs would be valuable as device interconnects and could create new opportunities for exploring novel 1D phenomena such as Luttinger liquids (*12-14*), plasmonics (*15-17*), charge density waves (*18, 19*), and superconductivity (*20, 21*). Thus far, the finite band gaps of GNRs synthesized using atomically-precise bottom-up fabrication techniques (*3, 4, 7, 9*) have been consistent with semiconducting theoretical predictions. New opportunities for achieving GNR metallicity arise from emergent topological concepts that allow placement of topologically-protected junction states at predetermined positions along the GNR backbone (22-26). These localized states each contribute a single unpaired electron at mid gap to the electronic structure (i.e., at $E = 0$) and so judicious placement of such zero-mode states raises the possibility of creating new metallic and magnetic configurations. Thus far, however, only semiconducting GNRs have been fabricated using this technique (*25, 26*).

Here we demonstrate a general approach for designing and fabricating metallic GNRs using the tools of atomically-precise bottom-up synthesis. This is accomplished by embedding localized zero-mode states in a symmetric superlattice along the backbone of an otherwise semiconducting GNR. Quantum mechanical hopping of electrons between the adjacent zero-mode states results in a metallic band as predicted by elementary tight-binding electronic structure models (*27*). Using scanning tunneling spectroscopy (STS) and first-principles theoretical modeling, we find that zero-modes confined to only one of graphene's two sublattices (i.e., sublattice-polarized states) result in narrow-band metallic phases that reside at the border of a magnetic instability. The metallic bandwidth of these GNRs, however, can be increased by more than a factor of 20 by intentionally breaking the GNR bipartite symmetry, thus resulting in robust metallicity. This is accomplished by inducing the formation of just two new carbon-carbon bonds per GNR unit cell (each unit cell contains 94 carbon atoms in the bottom-up synthesized GNRs presented here). This dramatic change in electronic structure from a seemingly minor chemical bond rearrangement arises from the loss of sublattice polarization that accompanies broken bipartite symmetry. This concept provides a useful new tool for controlling GNR metallicity and for tuning GNR electronic structure into new physical regimes.

Our strategy for designing metallic GNRs is based on a simple theorem: if a piece of graphene has a surplus of carbon atoms ($\Delta N$) on sublattice $A$ versus sublattice $B$, then this results in a minimum of $\Delta N = N_A - N_B$ eigenstates localized on the $A$ sublattice at $E = 0$ ("zero-modes") that are each occupied by one electron. Here $N_A$ ($N_B$) is the number of atoms residing on sublattice $A$ ($B$) (see Supplementary Materials for a derivation under nearest-neighbor interactions only). Expanding this idea to 1D GNR systems with a periodic sublattice imbalance, one can construct a low-energy effective tight-binding model to describe the resulting electronic bands by introducing a parameter, $t$, that represents electron hopping between adjacent zero-modes. This concept can be used to design metallic GNRs by providing them with a unit cell that contains a surplus of two carbon atoms on sublattice $A$ ($\Delta N = 2$). Under this construction there are two relevant hopping amplitudes, the intra-cell hopping amplitude ($t_1$) and the inter-cell hopping amplitude ($t_2$). A tight-binding analysis of this situation leads to the well-known Su-Schrieffer-Heeger (SSH) (*27*) dispersion relationship for the zero-mode bands:

$$E_{\pm}(k) = \pm\sqrt{|t_1|^2 + |t_2|^2 + 2|t_1||t_2|\cos(k+\delta)} \qquad (1)$$

where $\delta$ is the relative phase between $t_1$ and $t_2$ (which in general are complex). Two bands result here since there are two zero-mode states per unit cell and the energy gap between them is $\Delta E = 2||t_1| - |t_2||$. If the two hopping amplitudes are identical, $|t_1| = |t_2|$, then the energy gap is reduced to zero and the resulting 1D electronic structure should be *metallic*.

Using this idea as a guide for creating metallic GNRs, we designed the GNR precursor molecule **1** (Fig. 1A). A graphene honeycomb lattice superimposed onto this molecule reveals that under cyclodehydrogenation the methyl group carbon atom attached to the central tetracene (highlighted grey in Fig. 1A) will fuse and provide one surplus carbon atom on sublattice *A* over sublattice *B* per monomer. Previous step-growth polymerizations of structurally related molecules (*26*) suggest that the surface polymerization of **1** will place the central tetracene unit in an alternating pattern on either side of the GNR growth axis. If polymerization proceeds in a head-to-tail configuration then the resulting GNRs feature two additional carbon atoms on sublattice *A* per unit cell (Fig. 1A). Following cyclodehydrogenation the anticipated GNR structure is comprised of short zigzag edges and prominent cove regions (reminiscent of a saw blade) and will herein be referred to as the sawtooth-GNR (sGNR). Based on the symmetry of the sGNR unit cell one anticipates that the hopping amplitudes $t_1$ and $t_2$ will be equal (Fig. 1A, red arrows) resulting in a metallic band structure for the sGNR. A caveat to this approach is the limited control over head-to-tail surface polymerization since head-to-head and tail-to-tail polymerizations place the extra carbon atoms on opposite sublattices and are expected to lead to gapped semiconductors (Fig. S1).

Addition of (10-bromoanthracen-9-yl)lithium to a suspension of **3** followed by dehydration of crude diol precursor yielded the molecular precursor **1** for sGNRs (Fig. S2). Precursor **1** was then deposited onto a clean Au(111) surface in ultrahigh vacuum (UHV) using a Knudsen cell evaporator (Materials and Methods). Fig. 1A shows a representative STM image of two precursor molecules on Au(111). Step-growth polymerization of **1** was induced by heating the surface to 200 °C for 20 min, followed by a second annealing step at 300 °C for 20 min to complete the cyclodehydrogenation. A topographic STM image of a sGNR segment resulting from head-to-tail polymerization is depicted in Fig. 1B. Prominent periodic bright spots are observed at the locations of the cove regions due to the non-planar conformation induced by the superposition of hydrogen atoms (Fig. S3A). Bond-resolved STM (BRSTM) further corroborates the sGNR structure (Fig. 1B). A representative image showing the distribution of head-to-tail (Fig. 1A), head-to-head, and tail-to-tail (Fig. S1) segments in the sGNR is depicted in Fig. 1C.

Prolonged annealing of sGNRs at temperatures >300 °C induces a secondary cyclodehydrogenation along the cove regions that leads to the formation of five-membered rings along the edges of sGNRs (Fig. 1A). While at 300 °C this transformation remains a rare event (<30% of coves regions undergo the secondary cyclization), we were able to force the vast majority of cove regions to undergo cyclodehydrogenation by annealing to higher temperatures (~95% of cove regions undergo the secondary cyclization at 350 °C). The resulting GNRs will herein be referred to as 5-sawtooth-GNRs (5-sGNRs) (Figs. 1D, E). BRSTM imaging of 5-sGNRs unambiguously confirms the presence of five-membered rings along the cove edges of the GNRs (Fig. 1D). The absence of periodic bright spots observed in the topographic STM image of 5-sGNRs (previously attributed to the superposition of hydrogen atoms along the cove edges of sGNRs) further corroborates the structural assignment (Fig. S3B).

In order to experimentally determine GNR metallicity, the electronic structure of sGNRs was characterized using STM spectroscopy. Fig. 2A shows a typical d$I$/d$V$ point spectrum obtained on a sGNR (d$I$/d$V$ spectroscopy provides a measure of the local density of states (LDOS) located beneath the STM tip). Distinctive features associated with valence band (VB) and conduction band (CB) edges can be seen at $V = -1.07 \pm 0.03$ V (State 1) and $V = 1.36 \pm 0.03$ V (State 3), respectively. Most prominent, however, is the sharp peak in LDOS (State 2) that is centered near V = 0 ($E_F$) ($0.02 \pm 0.02$ V). This peak continuously spans energies both below and above $E_F$, a clear signature of a gapless, metallic density of states. d$I$/d$V$ imaging of the wavefunction of these metallic sGNR states shows a characteristic serpentine pattern that snakes back and forth across the sGNR width (Fig. 2B). The valence and conduction band edge states, in contrast, have their highest intensity along the armchair edges of the GNR (Fig. 2B), consistent with previous measurements of conventional semiconducting GNRs under similar conditions (*28, 29*).

A similar experimental analysis was also performed on the fused 5-sGNRs as depicted in Fig. 3A. The point spectroscopy of 5-sGNRs was seen to be quite different from that of sGNRs. While features associated with the valence band edge ($V = -1.12 \pm 0.03$ V, State 1) and conduction band edge ($V = 1.64 \pm 0.09$ V, State 3) of 5-sGNRs are observed at similar energies compared to sGNR states, the 5-sGNR spectrum does not feature a central peak at $V = 0$ (Fig. 3A). Instead it exhibits a shallow dip at $V = 0$ and a broad density of states (DOS) feature that spans an energy range above and below $E_F$. The electronic wavefunctions corresponding to States 1-3 in 5-sGNRs are similar to the corresponding features in sGNRs except for the lack of periodic bright spots associated with the non-planar cove edges (Fig. 3B). For example, d$I$/d$V$ images performed at biases near the conduction and valence band edges show the LDOS concentrated at the armchair edges while near $V = 0$ we observe a serpentine pattern (Fig. 3B) very similar to the metallic state seen in sGNRs. This state persists as the bias is swept across the dip at $V = 0$ over a wide energy range (–0.10 V < $V$ < 0.07 V) (Fig. S4). This implies that 5-sGNRs are also metallic and that the LDOS dip observed near $V = 0$ is not an energy gap but rather a metallic density-of-states feature.

We further explored the apparent metallicity of sGNRs by using *ab initio* density functional theory (DFT). Fig. 4C shows the resulting band structure calculated for an isolated sGNR using the local density approximation (LDA). Two narrow bands (denoted zero-mode bands (ZMBs)) are observed bracketing $E_F$, while CB and VB edges can be seen at much further energies. The two bands bracketing $E_F$ have no bandgap and are fit well by the SSH expression (Eq. 1) with $t_1 = t_2 = 5.2$ meV and $\delta = 0$ (Fig. 4C, red dashed lines), and are also stable against Peierls distortion (*30*) (as confirmed via supercell calculations). The resulting theoretical density of states (Fig. 2C) shows a single peak centered at $E_F$ as well as VB/CB peaks at lower/higher energies, in good agreement with the STM point spectroscopy for sGNRs (Fig. 2A). The theoretical wavefunction maps (Fig. 2D) also match the experimental d$I$/d$V$ maps obtained at $E_F$ and at the band edge energies, providing further evidence of metallicity in sGNRs.

While our sGNRs clearly match the metallic predictions of the symmetric SSH model, a potential complication is the very narrow metallic sGNR bandwidth (~21 meV). Metals with high DOS at $E_F$ are often unstable to Mott insulator transitions or magnetic phase transitions as dictated by the Stoner criterion (*31, 32*). The metallic behavior here may be due to the fact that spin polarization is not accounted for in our simplified tight-binding or LDA-based calculations. To test for this type of magnetic instability in sGNRs, we calculated the sGNR band structure using the local spin density approximation (LSDA) for an isolated sGNR. The result (Fig. S5) shows that the sGNR electronic structure does, in fact, undergo a ferromagnetic phase transition which opens a 200 meV energy gap about $E_F$. We show that the reason we do not see a gap experimentally is due to a combined effect of *p*-doping and surface electric fields induced by the underlying

Au(111) substrate. When these are properly accounted for in our DFT calculation, the gap does, indeed, vanish at the LSDA level and the metallic result is recovered (Fig. S5D). Therefore, while it is technically correct to say that sGNR/Au(111) is metallic, our DFT calculation predicts that a significant energy gap will open up and metallicity will be lost due to a magnetic phase transition as soon as this sGNR is removed from the Au(111) surface. While this represents an interesting and potentially useful 1D magnetic phase transition, the question remains whether it is possible to engineer a sGNR with more robust metallicity that would not suffer this fate.

This question can be answered by looking no further than the 5-sGNR whose metallic DOS features are much wider in energy than the narrow peak at $E_F$ seen for sGNRs (Figs. 2, 3). To clarify the robustness of 5-sGNR metallicity we also analyzed its electronic structure via *ab initio* DFT calculations. At the LDA level, the 5-sGNR band structure does, indeed, show a much wider metallic band than the corresponding sGNR band structure (Figs. 4C, D). When the SSH expression (i.e., the tight-binding result from Eq. 1) is fit to the 5-sGNR DFT-LDA band structure, we find a hopping amplitude of $t_1 = t_2 = 120$ meV, which corresponds to a bandwidth 23 times larger than the sGNR DFT-LDA bandwidth (Fig. 4D, red dashed lines). This is also reflected in the calculated DOS (Fig. 3C) which shows a broad U-shaped feature (with peaks characteristic of 1D van Hove singularities), consistent with the experimental dip in LDOS observed at $V = 0$ for 5-sGNRs (Fig. 3A). The theoretical LDOS patterns calculated for the 5-sGNR band edge and metallic states (Fig. 3D) also correspond well to the 5-sGNR experimental $dI/dV$ images (Fig. 3B). Our DFT calculations of the 5-sGNR at the LSDA level additionally show no signs of magnetism and are identical to the LDA-based results (Fig. S6). We conclude that 5-sGNRs exhibit robust metallicity with a much wider bandwidth than sGNRs, both experimentally and theoretically, and are *not* expected to undergo a magnetic phase transition upon transfer from Au(111) to an insulator.

The last question that we address is how the seemingly small structural difference between 5-sGNRs and sGNRs leads to such a large difference in their electronic behavior. The dramatic increase in bandwidth observed for 5-sGNRs can be understood as a result of the loss of sublattice polarization of the electron wavefunction induced via the disruption of the graphene bipartite lattice symmetry due to the new five-membered ring bonds (which bridge what were previously open coves). This can be understood by remembering that the two extra atoms added to the sGNR unit cell on sublattice $A$ result in two new $E = 0$ eigenstates (zero-modes) per unit cell whose wavefunctions are also confined to sublattice $A$. This sublattice polarization is preserved in the sGNR Bloch waves for the two in-gap bands (Fig. 5B), and the sGNR bandwidth is determined by the effective amplitude ($t_{eff}$) for an electron to hop between adjacent zero-modes (Fig. 5A). Because the zero-modes are all on the same sublattice it can be shown that $t_{eff} \propto t'$ where $t'$ is the second-nearest-neighbor hopping amplitude of graphene (since there is no zero-mode state density on sublattice $B$). In the case of 5-sGNRs, however, the bipartite lattice is *disrupted* by the bond that closes the coves; the zero-modes are thus no longer sublattice polarized (Fig. S7B). Consequently, the resulting Bloch waves are no longer sublattice polarized (i.e., both sublattices now exhibit state density (Fig. 5C)) and so $t_{eff} \propto t$ where $t$ is the nearest-neighbor hopping amplitude of graphene (Fig. 5A) (see SI for additional details). This explanation is consistent with the ratio of the bandwidths of the two GNRs (~23) which falls within the range of accepted values for $t/t'$ (*33*). The key insight here is that the loss of sublattice polarization (i.e., through intentional fusion of five-membered rings along the cove edges) greatly increases the effective overlap of adjacent localized zero-mode states and strongly enhances the metallic bandwidths. This provides a useful new design criterion for engineering robust metallic systems from zero-mode superlattices in carbon networks.

In conclusion, we have successfully demonstrated the ability to rationally design 1D metallic GNRs by embedding symmetric zero-mode superlattices into otherwise semiconducting GNR backbones. Our results provide a general strategy for introducing zero-modes into graphene-based materials, and also reveal the hidden role of sublattice polarization in controlling the emergent band structure of these systems. This general approach provides new opportunities for creating nanoscale electrical devices, and for exploring novel electronic and magnetic phenomena in a new class of 1D metallic systems.

**Acknowledgments**

Research supported by the Office of Naval Research MURI Program N00014-16-1-2921 (molecular design, STM spectroscopy, band structure), by the US Department of Energy (DOE), Office of Science, Basic Energy Sciences (BES) under the Nanomachine Program award number DE-AC02-05CH11231 (surface growth, image analysis), by the Center for Energy Efficient Electronics Science NSF Award 0939514 (precursor synthesis), and by the National Science Foundation under grants DMR-1508412 (LSDA simulation) and DMR-1839098 (zero-mode analysis). Computational resources have been provided by the DOE Lawrence Berkeley National Laboratory's NERSC facility and by the NSF through XSEDE resources at NICS.

D.J.R., G.V., J.J. , S.G.L., M.F.C. and F.R.F. initiated and conceived the research, G.V., R.M. and F.R.F. designed, synthesized and characterized the molecular precursors, D.J.R., C.B., T. Chen and M.F.C. performed on-surface synthesis and STM characterization and analysis, J.J., T.Cao. and S.G.L. performed the DFT calculations and the theoretical analysis that predicted and interpreted the STM data. All authors contributed to the scientific discussion.

The authors declare no competing interests.

All data are available in the main text or the Supplementary Materials.

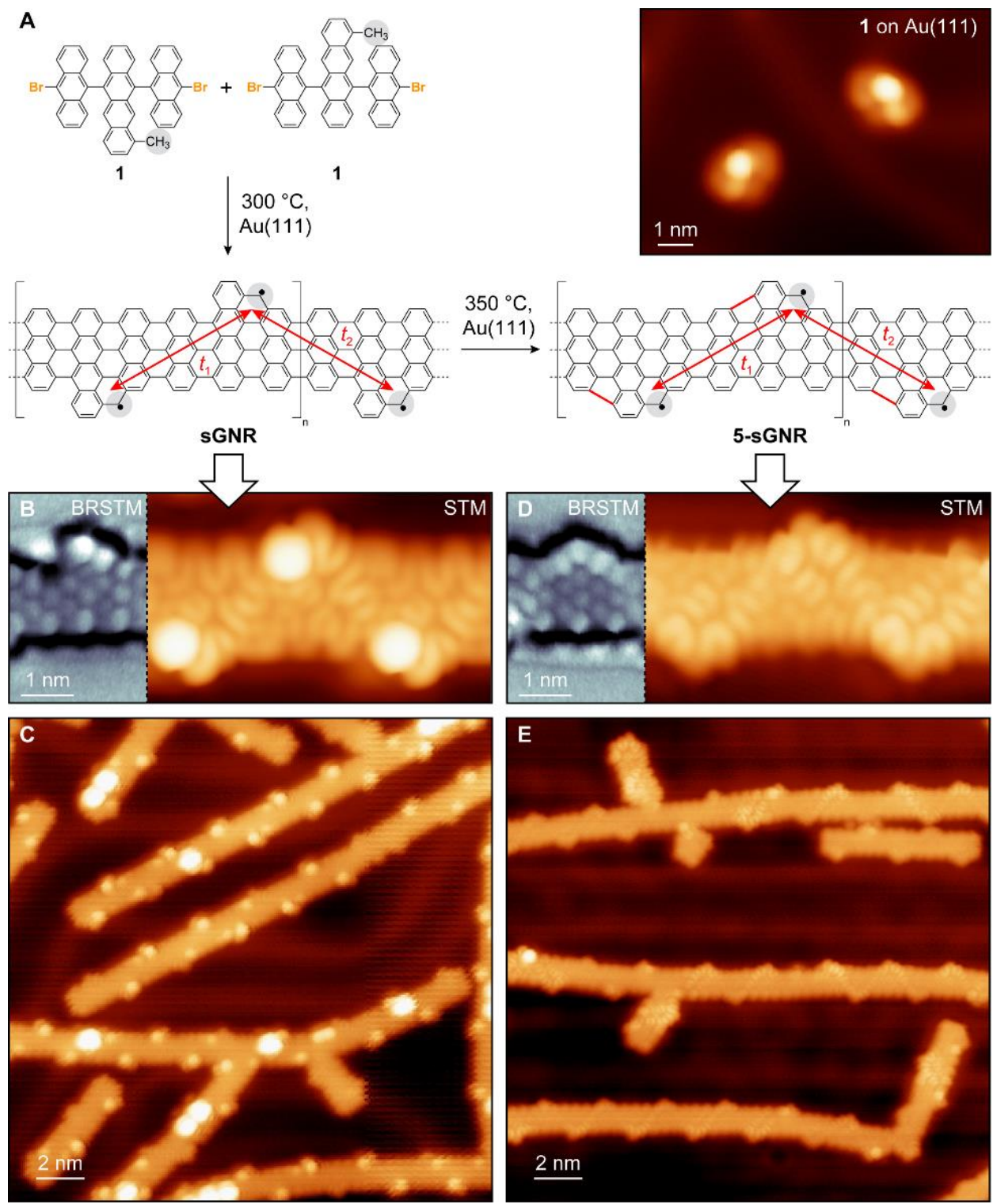

**Fig. 1.** Bottom-up synthesis of sawtooth GNRs. (**A**) Schematic representation of bottom-up growth of both sGNRs and 5-sGNRs from molecular precursor **1**. Inset: STM topograph of two isolated

monomers of **1** deposited on Au(111) ($I_t$ = 30 pA, $V_s$ = 1.000 V). (**B**) STM topograph of a segment of a sGNR ($I_t$ = 80 pA, $V_s$ = 0.006 V). Inset shows a bond-resolved STM (BRSTM) image of a cyclized monomer unit with an intact cove region ($I_t$ = 110 pA, $V_s$ = 0.010 V, $V_{AC}$ = 10 mV). (**C**) Larger-scale image of sGNRs ($I_t$ = 30 pA, $V_s$ = − 1.100 V). (**D**) STM topograph of a segment of a 5-sGNR ($I_t$ = 1.5 nA, $V_s$ = − 0.100 V). BRSTM image in inset shows how cove regions fuse to form five-membered rings ($I_t$ = 110 pA, $V_s$ = 0.010 V, $V_{AC}$ = 10 mV). (**E**) Large-scale image of 5-sGNRs ($I_t$ = 20 pA, $V_s$ = 0.010 V).

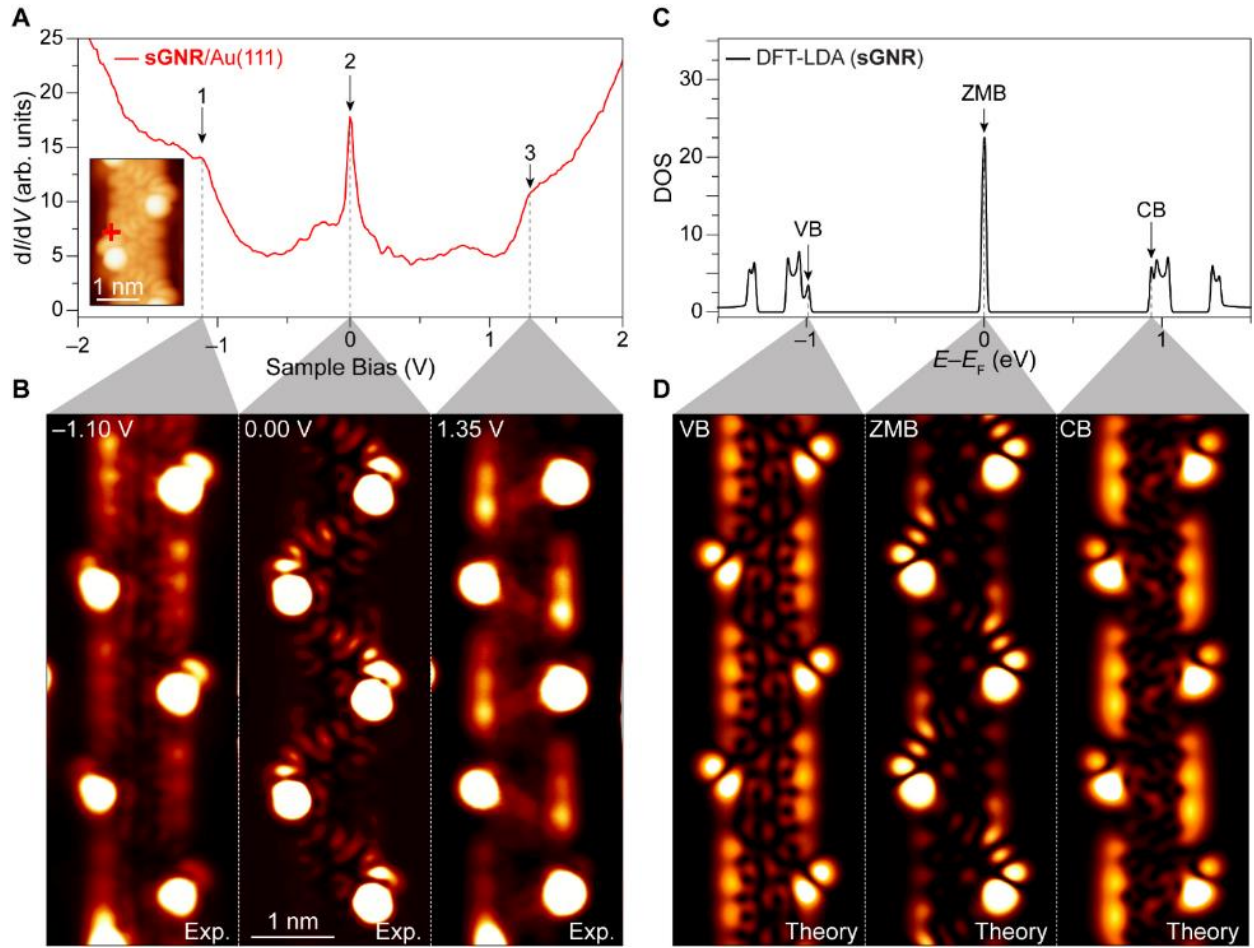

**Fig. 2.** Electronic structure of sGNRs. (**A**) d$I$/d$V$ point spectroscopy of a sGNR/Au(111) performed at the position shown in the inset (spectroscopy parameter: $V_{AC}$ = 10 mV. Imaging parameters: $I_t$ = 80 pA, $V_s$ = 0.006 V). (**B**) Constant-height d$I$/d$V$ maps of sGNRs conducted at the biases indicated in (A) (spectroscopy parameters: $V_{AC}$ = 20 mV for States 1 and 3, $V_{AC}$ = 4 mV for State 2). Constant-height d$I$/d$V$ maps were subjected to background subtraction of substrate LDOS as described in Fig. S8 (*34*). (**C**) DFT-LDA calculated DOS of the sGNR (spectrum broadened by 10 meV Gaussian). Van Hove singularities near $E - E_F = 0$ are merged because of gaussian smearing. (**D**) DFT-calculated LDOS of an sGNR at energies shown in (C) (LDOS sampled at a height of 3.5 Å above the plane of the sGNR).

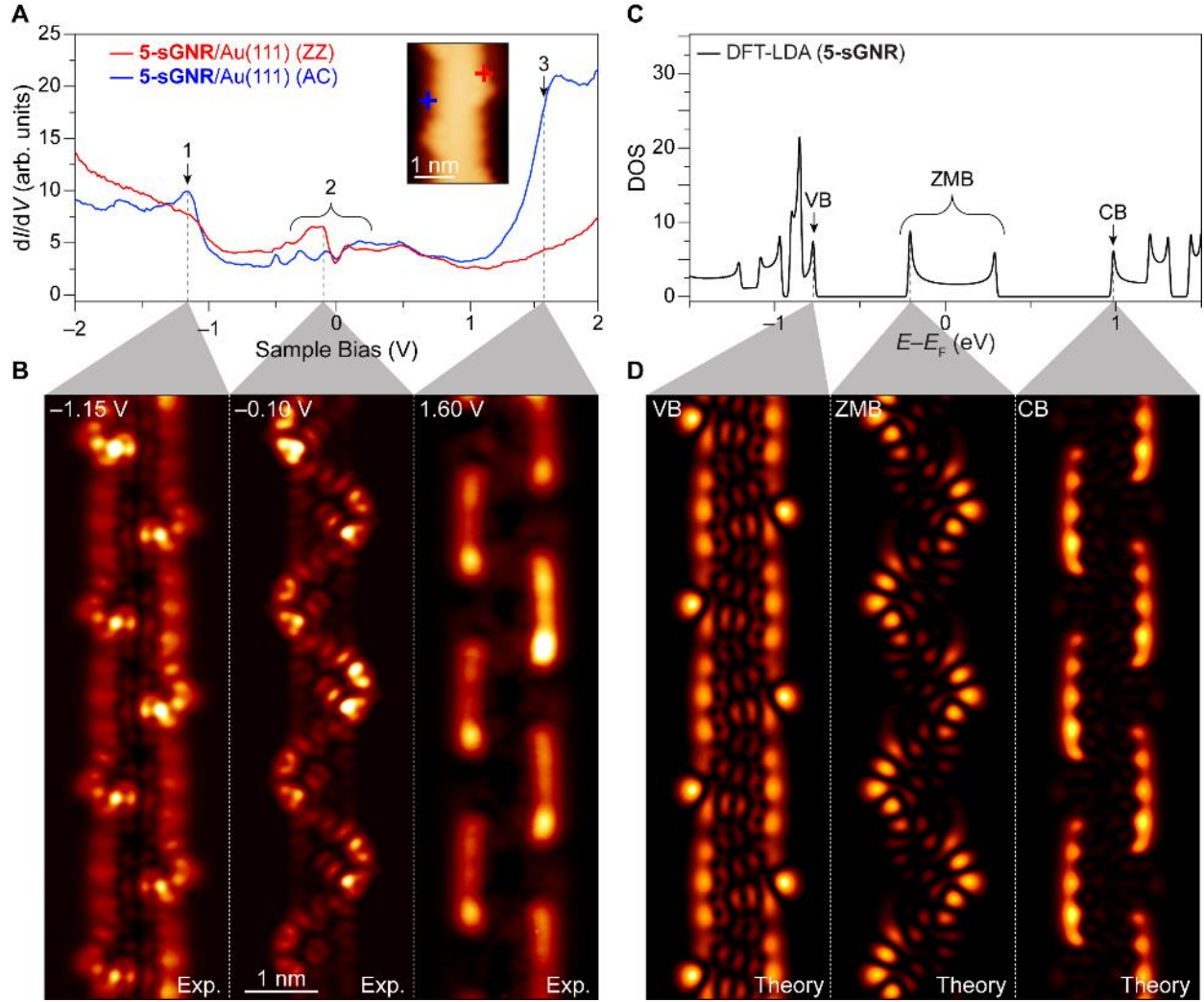

**Fig. 3.** Electronic structure of 5-sGNRs. (**A**) d$I$/d$V$ point spectroscopy conducted on 5-sGNR/Au(111) at the armchair (blue) and zigzag (red) positions marked in the inset (spectroscopy parameter: $V_{AC}$ = 10 mV. Imaging parameters: $I_t$ = 60 pA, $V_s$ = – 0.100 V). (**B**) Constant-height d$I$/d$V$ maps of 5-sGNRs conducted at the biases indicated in (A) (spectroscopy parameter: $V_{AC}$ = 20 mV). Constant-height d$I$/d$V$ maps were subjected to background subtraction of substrate LDOS as described in Fig. S8 (*34*). (**C**) DFT-LDA calculated DOS of the 5-sGNR (spectrum broadened by 10 meV Gaussian). (**D**) DFT-LDA calculated LDOS of a 5-sGNR at energies shown in (C) (LDOS sampled at a height of 3.5 Å above the plane of the 5-sGNR).

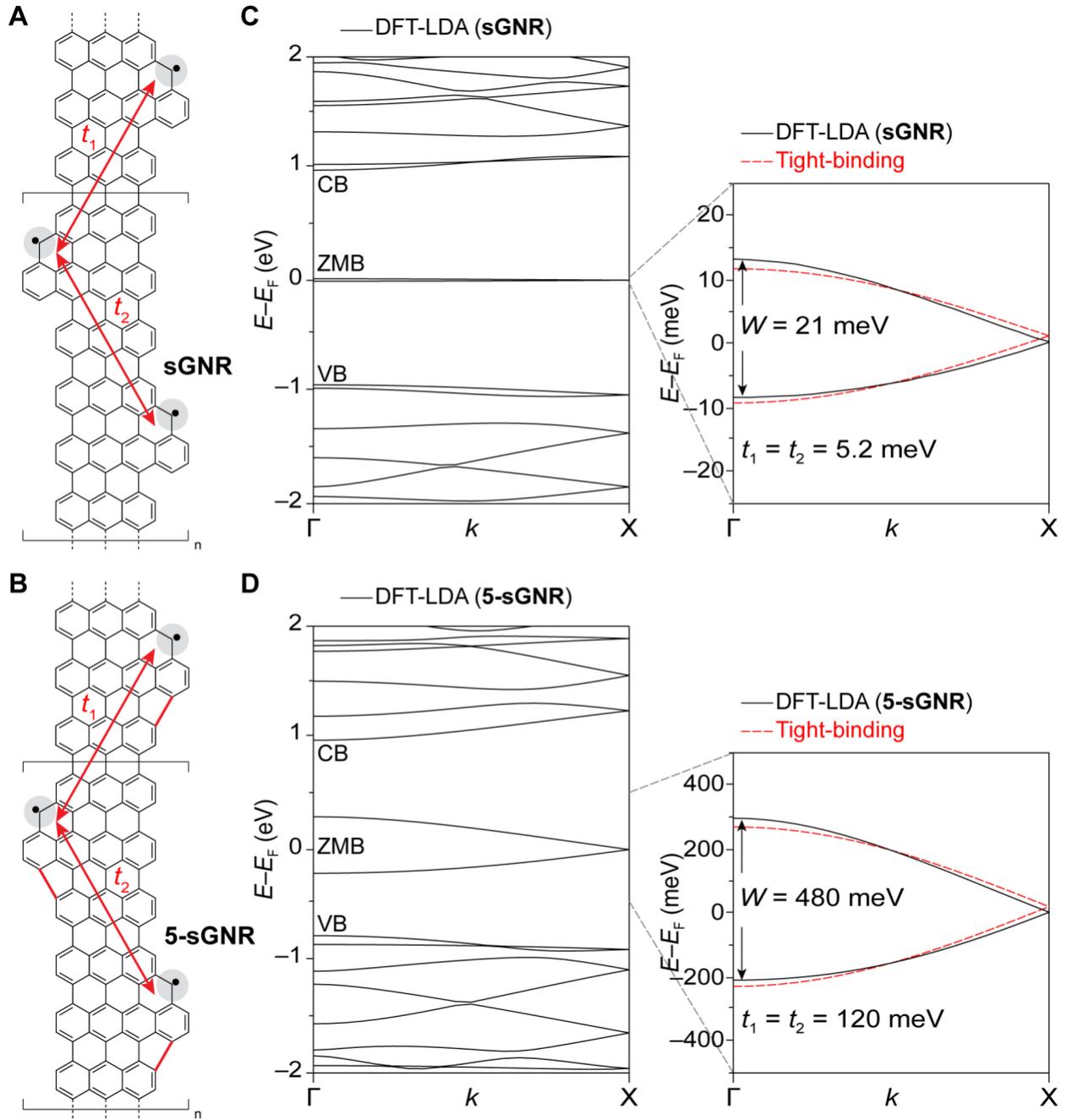

**Fig. 4.** Zero-mode band structure. Schematic representation of inter- and intracell hopping between localized zero-modes embedded in (**A**) sGNRs and (**B**) 5-sGNRs. (**C**) Left panel: DFT-LDA calculated band structure for sGNRs. Frontier bands are labelled VB, ZMB, and CB. Right panel: tight-binding fit (red) to DFT-LDA band structure yields hopping parameter $t_1 = t_2 = 5.2$ meV.

(**D**) The same as (C) but for 5-sGNRs. Hopping parameter for 5-sGNR (and corresponding bandwidth) is 23 times larger than for sGNR.

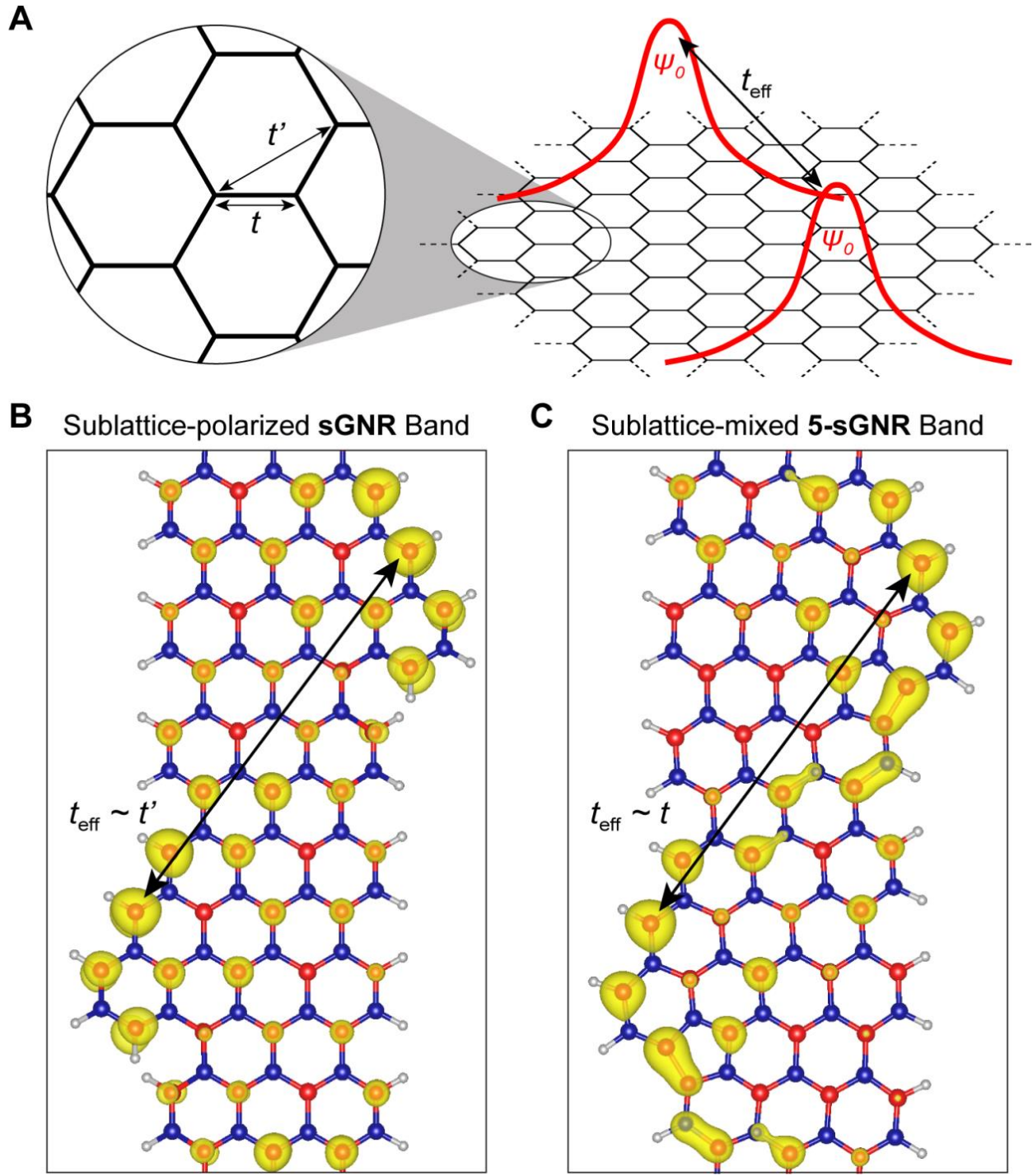

**Fig. 5.** Zero-mode engineering in GNRs. (**A**) Diagram of effective hopping $t_{\text{eff}}$ between two localized states (labeled $\psi_0$) embedded in graphene. Inset: schematic representation of the first ($t$) and second ($t'$) nearest-neighbor hopping parameters of graphene. (**B**) DFT-calculated

wavefunction isosurface of a sGNR for states near $E = 0$ (5% charge density isosurface shown). (**C**) Same as (B) but for 5-sGNRs. Different sublattices are denoted with different colors (*A* sublattice in red and *B* sublattice in blue). The sGNR wavefunction is completely sublattice polarized, while the 5-sGNR wavefunction is sublattice mixed and more delocalized.